\documentclass{article} % For LaTeX2e
\usepackage{iclr2026_conference,times}
\iclrfinalcopy 

\usepackage{amsmath,amsfonts,bm}

\def\eqref#1{equation~\ref{#1}}
\def\1{\bm{1}}

\DeclareMathAlphabet{\mathsfit}{\encodingdefault}{\sfdefault}{m}{sl}
\SetMathAlphabet{\mathsfit}{bold}{\encodingdefault}{\sfdefault}{bx}{n}

\usepackage{hyperref}
\definecolor{black}{rgb}{0.0,0.0,0.0}
\definecolor{DeepDarkBlue}{rgb}{0.0745, 0.1490, 0.6314}
\hypersetup{colorlinks,breaklinks,
            linkcolor=DeepDarkBlue,urlcolor=DeepDarkBlue,
            anchorcolor=DeepDarkBlue,citecolor=DeepDarkBlue}
\usepackage{url}

\usepackage{lipsum}
\usepackage{array}
\usepackage{tabularx}
\usepackage{makecell}
\usepackage{multirow}
\usepackage{booktabs}

 \usepackage{graphicx}
\usepackage{subcaption}
\usepackage{fontawesome5}

\title{Developing an ESG-Oriented Large Language Model through ESG Practices}

\author{\textbf{Gabriel Assis}$^{\clubsuit}$~\faEnvelope~\thanks{Corresponding author.} \quad
\textbf{\& Ayrton Surica}$^{\clubsuit}$ \quad
\textbf{\& Pedro Kroll}$^{\clubsuit}$ \quad
\textbf{\& Gabriela Aires}$^{\diamondsuit}$\quad \\
\textbf{\& Darian Rabbani}$^{\diamondsuit}$\quad 
\textbf{\& Edson Bollis}$^{\diamondsuit}$\quad
\textbf{\& Lucas Pellicer}$^{\diamondsuit}$\quad
\textbf{\& Aline Paes}$^{\clubsuit}$~\faEnvelope\\[2pt]
$^{\clubsuit}$Universidade Federal Fluminense, Instituto de Computação, Niterói, RJ, Brazil\\
$^{\diamondsuit}$Instituto de Ciência e Tecnologia Itaú, São Paulo, SP, Brazil\\[2pt]
\texttt{assisgabriel@id.uff.br}\quad \texttt{alinepaes@ic.uff.br}
}

\begin{document}

\maketitle

\begin{abstract}

Environmental, Social, and Governance (ESG) considerations play a central role in contemporary financial decision-making. In parallel, Large Language Model (LLM) applications in this domain have primarily emphasized well-defined discriminative tasks, such as classification or scoring, which have proven effective for structured analysis and benchmarking. However, this prevailing focus offers limited support for more interactive and generative ESG scenarios, where embedded domain knowledge and contextual understanding are essential. In this work, we propose an ESG-oriented adaptation pipeline for LLMs that integrates ESG principles not only as a target domain, but also as guiding constraints throughout training and evaluation. Building on the Qwen-3-4B architecture, we explore parameter-efficient adaptation strategies using Low-Rank Adaptation (LoRA) and the Instruction-Residual Method (IRM) to produce three ESG-specialized models. We evaluate the proposed models on ESG question answering under both zero-shot and knowledge-augmented settings, using a diverse set of generative, semantic, readability, and environmental impact metrics. Our results show that the ESG-adapted models consistently outperform their original counterparts and competitive baselines such as Llama-3 and Gemma-3. Although limitations remain in tool-based knowledge integration, this work establishes a foundation for ESG-oriented language generation and highlights the importance of responsible, domain-aware LLM adaptation.

\end{abstract}

\section{Introduction}\label{sec:intro}

Environmental, Social, and Governance (ESG) considerations have become a critical pillar of sustainable development, corporate accountability, and financial decision-making~\citep{zhang2025mmesgbench,he-etal-2025-esgenius,esgbertshort}. Rather than addressing isolated concerns, ESG provides an integrated framework for evaluating how organizations manage environmental impact, social responsibility, and governance practices within their strategic operations~\citep{esgbertshort}.

Once viewed primarily as compliance or reporting mechanisms, ESG practices are now widely recognized as drivers of long-term financial performance. Prior work shows that strong ESG alignment is associated with reduced capital costs, improved stakeholder trust, and increased innovation capacity~\citep{friede2015esg}. As a result, ESG has evolved into a strategic signal of resilience and competitiveness, increasingly shaping investment strategies, regulatory oversight, and corporate planning, while influencing how organizations assess risk, allocate capital, and define long-term value creation.

In parallel, recent advances in artificial intelligence, particularly Large Language Models (LLMs), have also begun to transform financial analysis and decision support~\citep{assis-icaif,esgbertshort,he-etal-2025-esgenius,esg-europe}. LLMs have demonstrated strong performance across tasks such as financial question answering~\citep{assis-icaif}, market analysis~\citep{wu2023bloomberggptlargelanguagemodel}, and trading-related applications~\citep{bhatia-etal-2024-fintral}. Within ESG contexts, however, their use has largely focused on narrow downstream tasks, including classification, scoring, or information extraction~\citep{esgbertshort,esg-europe}. While effective, these applications do not fully leverage LLMs' generative capabilities, nor do they provide models with embedded ESG knowledge suitable for interactive reasoning or open-ended generation.

This gap motivates the development of ESG-oriented generative language models. In this work, we propose an ESG-aware methodology for adapting LLMs that explicitly integrates ESG principles not only as a target domain, but also as part of the training process itself. We build upon the Qwen-3-4B~\citep{yang2025qwen3technicalreport} architecture and explore parameter-efficient adaptation techniques, including Low-Rank Adaptation (LoRA)~\citep{hu2022lora} and the Instruction-Residual Method (IRM)~\citep{jindal2024balancingcontinuouspretraininginstruction}, to specialize the model for ESG applications.

Our approach emphasizes multiple ESG dimensions throughout the pipeline: governance, through the use of curated and openly available ESG data; environmental responsibility, by monitoring and minimizing training-related carbon impact; and social considerations, by evaluating the readability and accessibility of generated outputs. We introduce three ESG-specialized models\footnote{\url{https://huggingface.co/collections/melll-uff/esg-models}} --- Qwen-3-4B-Base–ESG, Qwen-3-4B-Instruct–ESG, and Qwen-3-4B-IRM–ESG --- and evaluate them against their original counterparts as well as strong baselines such as Llama-3~\citep{llama3} and Gemma-3~\citep{gemmateam2025gemma3technicalreport}. Our results demonstrate consistent improvements and strong zero-shot performance, positioning these models as a foundation for future ESG-oriented language generation.

The remainder of this paper is organized as follows. Section~\ref{sec:rw} reviews related work, Section~\ref{sec:found} presents the machine learning foundations underlying model adaptation, Section~\ref{sec:method} describes the training and evaluation methodology, Section~\ref{sec:results} reports and discusses the results, and Section~\ref{sec:remarks} concludes the paper.

\section{Related Work}\label{sec:rw}

Financial-oriented LLMs have emerged since their first prominent representative, BloombergGPT~\citep{wu2023bloomberggptlargelanguagemodel}. More recent approaches focus on further improving and continuously updating general financial language models~\citep{bhatia-etal-2024-fintral}, as well as on specialization across multiple languages or specific downstream tasks, such as sentiment analysis and ESG-related applications, including classification and scoring~\citep{FinGEITje,esgbertshort,finllama,esg-europe}.

In this landscape, FinTral~\citep{bhatia-etal-2024-fintral} stands out as a notable model. Built upon the open Mistral-7B~\citep{jiang2023mistral7b} architecture, FinTral is specialized for financial applications and reports performance comparable to GPT-4o~\citep{openai2024gpt4ocard} on several financial tasks, including summarization, credit prediction, and stock movement prediction. Similarly based on Mistral-7B, FinGEITje~\citep{FinGEITje} targets banking-related tasks not only in English but also introduces specialization for Dutch, representing one of the first initiatives to move beyond English-centric financial LLMs. FinLlama~\citep{finllama}, built on Llama-2-7B~\citep{touvron2023llama2openfoundation}, is specifically fine-tuned for sentiment analysis and has been incorporated into trading-oriented frameworks. Beyond generative architectures, encoder-based approaches have also been explored in the financial domain. In particular, the ESG-BERT~\citep{esgbertshort} family builds upon the line of work initiated by models such as FinBERT~\citep{araci2019finbertfinancialsentimentanalysis}, extending financial encoder-based architectures to ESG-focused classification tasks. ESG-BERT comprises a set of classifiers that predict scores for textual evidence across each ESG pillar.

In addition to proposing specialized financial LLMs, several studies investigate the application of general-purpose models --- such as the aforementioned Mistral and Llama --- to financial tasks, with some works explicitly addressing environmental impact and ESG-related dimensions. For instance, \citet{assis-icaif} evaluate a broad range of LLMs and LLM-based agents on a financial question answering task, highlighting the strong performance of the Qwen~\citep{qwen2025qwen25technicalreport,yang2025qwen3technicalreport} family. Importantly, their study also provides a comprehensive assessment of execution time, energy consumption, and carbon emissions. Similarly, \citet{he-etal-2025-esgenius} conduct an extensive evaluation of 50 LLMs in ESG-related tasks, also reporting a Qwen-family model among the top-performing open architectures. Furthermore, \citet{esg-europe} analyze how fine-tuning general-purpose LLMs affects ESG Activity Detection, demonstrating that task-specific adaptation yields consistent improvements, while also maintaining transparency regarding experimental setup and computational costs.

Overall, the works in this section demonstrate substantial progress in financial and ESG-oriented language modeling. However, to the best of our knowledge, there is still no generative-purpose foundation model explicitly designed for ESG contexts, capable of supporting interactive use cases or broader tasks such as open-ended question answering. This work contributes toward filling this gap by proposing an ESG-oriented generative model, developed with emphasis on transparency, computational efficiency, and responsible practices aligned with ESG principles.

\section{Theoretical Foundations}\label{sec:found}

Since the establishment of the \emph{foundation model}~\citep{bommasani2022opportunitiesrisksfoundationmodels} paradigm, it has become common practice to reuse large pretrained models and further specialize them for downstream domains or tasks through \emph{Supervised Fine-Tuning} (SFT)~\citep{NEURIPS2022_b1efde53}. In this process, pretrained parameters serve as an initialization, and the model weights are updated by minimizing a supervised objective, typically resulting in a substantially lower computational cost and reduced data requirements than training a model entirely from scratch.

Formally, given a pretrained model with parameters $\theta$ and a supervised dataset $\mathcal{D} = \{(x_i, y_i)\}_{i=1}^{N}$, SFT optimizes the following objective:
\[
\mathcal{L}_{\mathrm{SFT}}(\theta) =
\mathbb{E}_{(x,y)\sim\mathcal{D}}
\left[
\ell\big(f_{\theta}(x), y\big)
\right],
\]
where $f_{\theta}$ denotes the language model, $y$ is usually the next token, and $\ell(\cdot)$ is typically the cross-entropy loss for next-token prediction.

SFT constitutes a central component of the modern LLM training stack, which can be broadly divided into three stages: \emph{pre-training}, \emph{mid-training}, and \emph{post-training}~\citep{olmo2025olmo3}. During pre-training, models are exposed to massive-scale corpora and trained autoregressively to acquire general linguistic competence and world knowledge. In the mid-training stage, models are further refined --- often still via SFT --- on more specialized datasets, such as domain-specific or preference-aligned corpora. Finally, post-training commonly relies on reinforcement learning–based techniques to induce reasoning, alignment, and instruction-following capabilities.

Despite its effectiveness, SFT can still require substantial computational resources, particularly as the parameter count of modern LLMs continues to scale. In this context, \emph{Low-Rank Adaptation} (LoRA)~\citep{hu2022lora} has emerged as a prominent parameter-efficient alternative~\citep{lin2025continuallearningsparsememory}. Instead of updating the full parameter set, LoRA injects trainable low-rank matrices into selected weight matrices while keeping the original parameters frozen.

Given a weight matrix $W \in \mathbb{R}^{d \times k}$, LoRA reparameterizes the update as:
\[
W' = W + \Delta W,
\quad
\text{where}
\quad
\Delta W = BA,
\]
with $A \in \mathbb{R}^{r \times k}$, $B \in \mathbb{R}^{d \times r}$, and $r \ll \min(d,k)$, substantially reducing computational costs.

Beyond efficiency gains, LoRA has also been shown to mitigate \emph{catastrophic forgetting}~\citep{MCCLOSKEY1989109, kotha2024understanding}, as its localized updates tend to interfere less with previously learned representations. This property makes it attractive for continual training scenarios, although LoRA-based adaptations are not entirely immune to forgetting~\citep{bidermanlora}.

To further address this limitation, techniques such as the \emph{Instruction-Residual Method} (IRM)~\citep{jindal2024balancingcontinuouspretraininginstruction}, which is based on the insight that updating model parameters prior to alignment stages can lead to more robust knowledge adaptation, emerge. IRM applies updates to the \emph{Base} version of a model and subsequently reconstructs an instruction-capable model by reintroducing alignment knowledge as a residual as follows.

Let $\theta_{\text{base}}$ denote the original base model parameters, $\theta_{\text{inst}}$ the instruction-tuned parameters, and $\theta_{\text{base}}'$ the adapted base model. IRM computes the instruction residual as:
\[
\Delta_{\text{inst}} = \theta_{\text{inst}} - \theta_{\text{base}},
\]
and constructs the final model as:
\[
\theta_{\text{final}} = \theta_{\text{base}}' + \Delta_{\text{inst}}.
\]

In this work, we investigate LoRA and IRM as complementary strategies for developing an ESG-specialized language model, aiming to balance computational efficiency with knowledge preservation while maintaining instruction-following capabilities.

\section{Training Methodology for an ESG-LLM}\label{sec:method}

This section describes the methodology adopted to develop and evaluate the ESG-oriented language models. We detail the datasets, training regimes, and adaptation techniques employed, as well as the evaluation protocol used to assess performance, efficiency, and environmental impact.

\subsection{Data}

The ESG-oriented model was trained and evaluated using the \textbf{ESG-QA dataset}~\citep{esgqa2025}. ESG-QA is a question answering benchmark comprising 87,261 question–answer–context triplets spanning the three ESG pillars, as illustrated in Table~\ref{tab:qa_examples}. The dataset is stratified by ESG pillar and split into 70\% / 10\% / 20\% partitions for training, validation, and testing, respectively. ESG-QA was selected due to its provenance from the ESG-BERT~\citep{esgbertshort} corpus, which was also used to pretrain the ESG-BERT model. This corpus comprises a diverse collection of ESG-related documents, including reports, articles, and other publicly available materials, offering rich, curated coverage of ESG content. As such, ESG-QA provides a high-quality, domain-consistent resource for training and evaluating ESG-focused models. 

\begin{table*}[htbp]
\centering
\scriptsize
\caption{Example QA-Context triplets by ESG pillar.}
\setlength{\tabcolsep}{2pt}
\renewcommand{\arraystretch}{1.1}
\newcolumntype{Y}{>{\raggedright\arraybackslash}X}
\begin{tabularx}{\textwidth}{@{}lYYY@{}}
\midrule
\textbf{Pillar} & \textbf{Question} & \textbf{Answer} & \textbf{Context} \\
\hline
Env. &
\textit{What specific emissions from waste management are included in the corporate carbon footprint according to the GHG Protocol Guidance?} &
\textit{Emissions included are those from CO$_2$ and biogenic carbon in waste, as well as CH$_4$ from the decomposition of biogenic materials in landfills or waste\mbox{-}to\mbox{-}energy (WTE) technologies.} &
\textit{Following GHG Protocol Guidance, CO$_2$ and biogenic carbon contained in waste and CH$_4$ emissions from the decomposition of biogenic materials in landfills or waste\mbox{-}to\mbox{-}energy (WTE) technologies are captured in the overall corporate carbon footprint.} \\

\hline

Soc. &
\textit{What primary risk factors make homeless youth more vulnerable to sex and labor trafficking?} &
\textit{Homeless youth are more vulnerable to sex and labor trafficking due to poverty, unemployment, a history of sexual abuse, and a history of mental health issues.} &
\textit{Homeless youth are vulnerable to both sex and labor trafficking because they tend to experience a higher rate of the primary risk factors to trafficking: poverty, unemployment, a history of sexual abuse, and a history of mental health issues.} \\

\hline

Gov. &
\textit{What are the key principles of sustainability that Trinity recognizes as critical to its long\mbox{-}term value?} &
\textit{The key principles of sustainability recognized by Trinity include environmental stewardship, safety and quality assurance, corporate social responsibility, governance, and diversity and inclusion.} &
\textit{We recognize that further integrating the key principles of sustainability --- including actions and transparency in environmental stewardship, safety and quality assurance, corporate social responsibility, governance, and diversity and inclusion --- is critical to enhancing Trinity’s long\mbox{-}term value.} \\
\hline
\end{tabularx}
\label{tab:qa_examples}
\end{table*}

Moreover, formulating ESG-QA in a question-answering format enables the evaluation of models in a conversational, interactive setting, aligning with the objectives of this research and the intended downstream use cases of ESG-oriented LLMs. Accordingly, models in this work were trained on context passages from the training split, while the validation set was used for model selection and configuration. The test split was held out and exclusively employed for final evaluations.

\subsection{Base Model}

The \textbf{Qwen-3 model}~\citep{yang2025qwen3technicalreport} was selected as the base architecture due to the Qwen family well-documented strong performance in financial and ESG-related tasks~\citep{assis-icaif,he-etal-2025-esgenius}. In addition, Qwen models have demonstrated competitive results on general-purpose benchmarks and are widely adopted within the AI ecosystem, making them a compatible choice within the broader modern LLM stack~\citep{artificialanalysis_qwen3}.

Specifically, we adopt the 4B-parameter variant, balancing empirical performance and computational efficiency. This model size enables effective specialization while remaining feasible under realistic resource constraints, aligning with our efficiency and sustainability objectives.

\subsection{Training Regimes}

\begin{figure*}[!ht]
  \centering
  \begin{subfigure}[t]{0.33\textwidth}
    \centering
    \includegraphics[width=\linewidth]{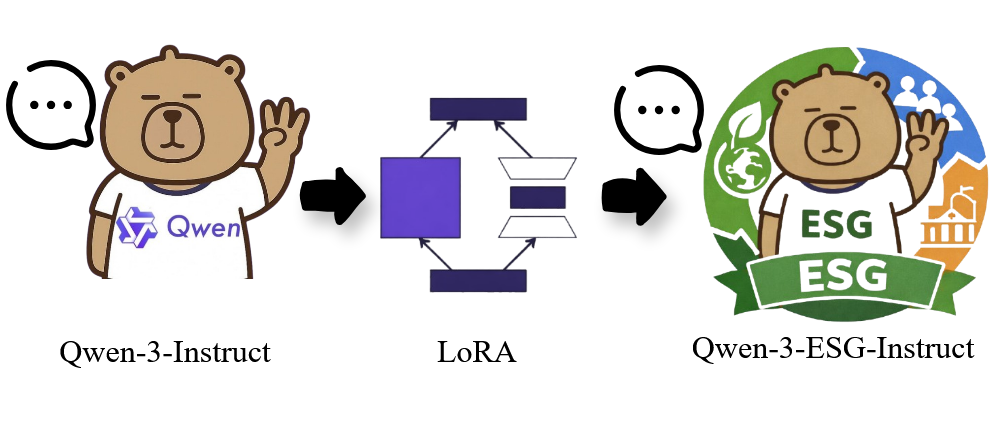}
    \caption{\textsc{LoRA} applied to Qwen-Instruct.}
    \label{fig:lora-qwen-instruct}
  \end{subfigure}
  \hspace{0.06\textwidth}
  \begin{subfigure}[t]{0.54\textwidth}
    \centering
    \includegraphics[width=\linewidth]{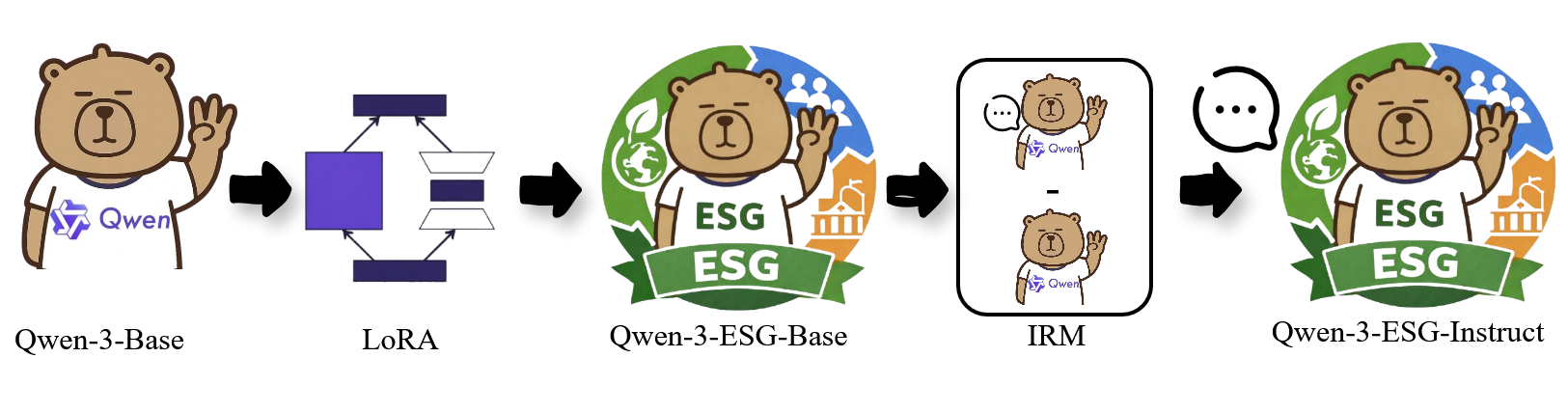}
    \caption{\textsc{IRM} starting from Qwen-Base.}
    \label{fig:irm-qwen-base}
  \end{subfigure}
  \caption{Training regimes applied for adapting the model to ESG context.}
  \label{fig:lora-irm}
\end{figure*}

Two training regimes\footnote{All training was conducted using Hugging Face Transformers~\citep{wolf-etal-2020-transformers} in Rio de Janeiro, relevant to CO$_2$ reporting.} were applied to adapt the Qwen-3-4B models to ESG contexts, as illustrated in Figure~\ref{fig:lora-irm}. In the first regime (Figure~\ref{fig:lora-qwen-instruct}), the instruction-tuned \emph{Qwen-3-4B-Instruct} variant is directly adapted using LoRA, in which weight updates are parameterized through a low-rank decomposition that introduces a small number of trainable matrices while keeping the original parameters frozen. The resulting low-rank update is then merged back into the original weights, incorporating the learned ESG representations into the model. Recent Qwen-3-4B releases also provide a separate \emph{thinking} variant in addition to the instruct model; however, in our validation experiments, this variant yielded inferior performance (see Section~\ref{sub:results}). In the second regime (Figure~\ref{fig:irm-qwen-base}), the same LoRA-based adaptation is applied, however, to the \emph{Qwen-3-4B-Base} variant, followed by the application of the IRM to recover instruction-following capabilities. This design choice is motivated by prior evidence that directly modifying instruction-tuned models can degrade existing capabilities, whereas IRM offers a mechanism to mitigate such effects by decoupling domain adaptation from alignment~\citep{jindal2024balancingcontinuouspretraininginstruction,lin2025continuallearningsparsememory}. These regimes enable us to analyze robustness and knowledge preservation in ESG-oriented adaptation.~Nevertheless, LoRA-based tuning improves efficiency and reduces computational overhead.

\begin{figure}[!ht]
\centering
  %\begin{flushright} % left-align within ICLR template
    \begin{subfigure}[t]{0.25\textwidth}
      \centering
      \includegraphics[width=\linewidth]{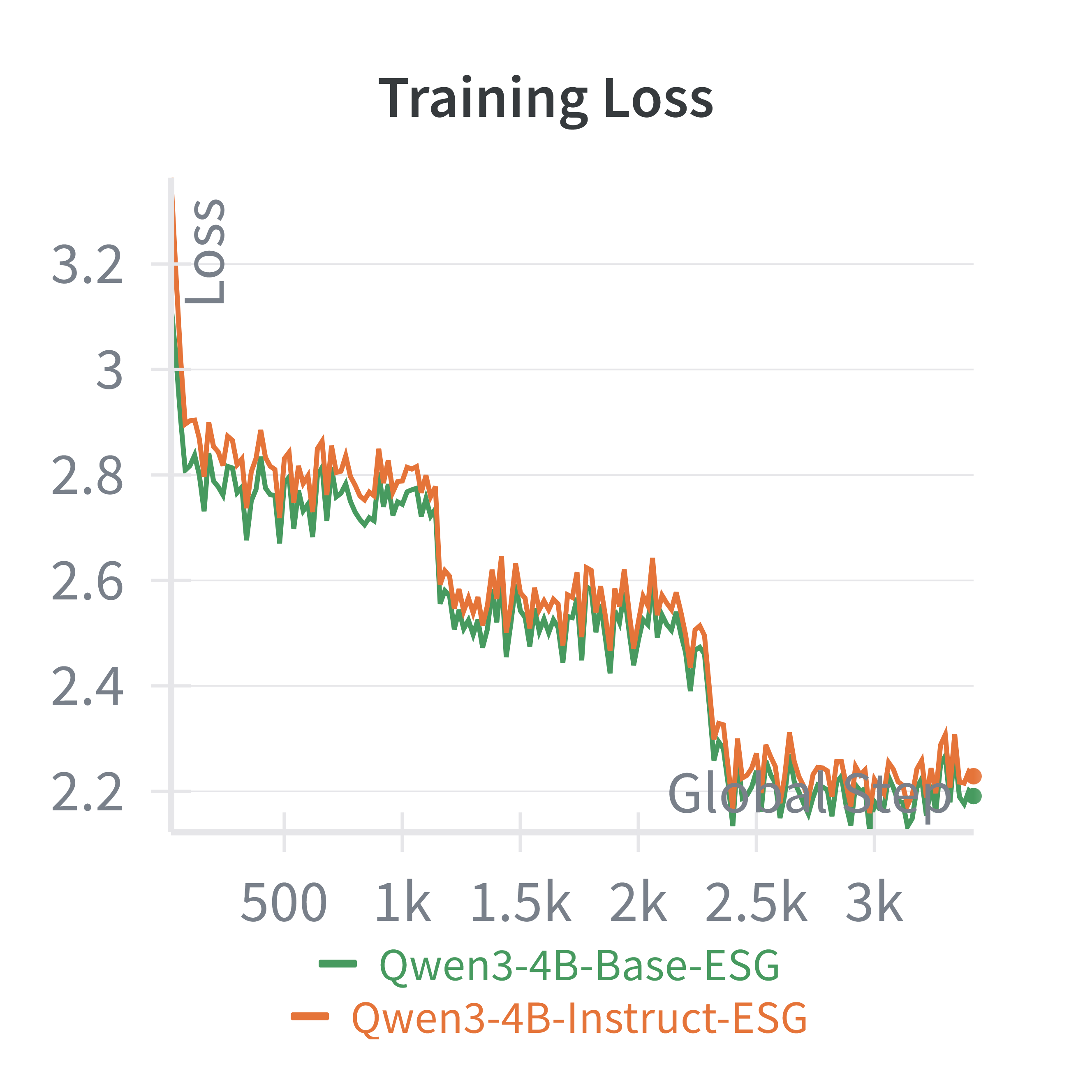}
      \caption{Training loss.}
      \label{fig:top}
    \end{subfigure}%\\[0.6em]
    \begin{subfigure}[t]{0.25\textwidth}
      \centering
      \includegraphics[width=\linewidth]{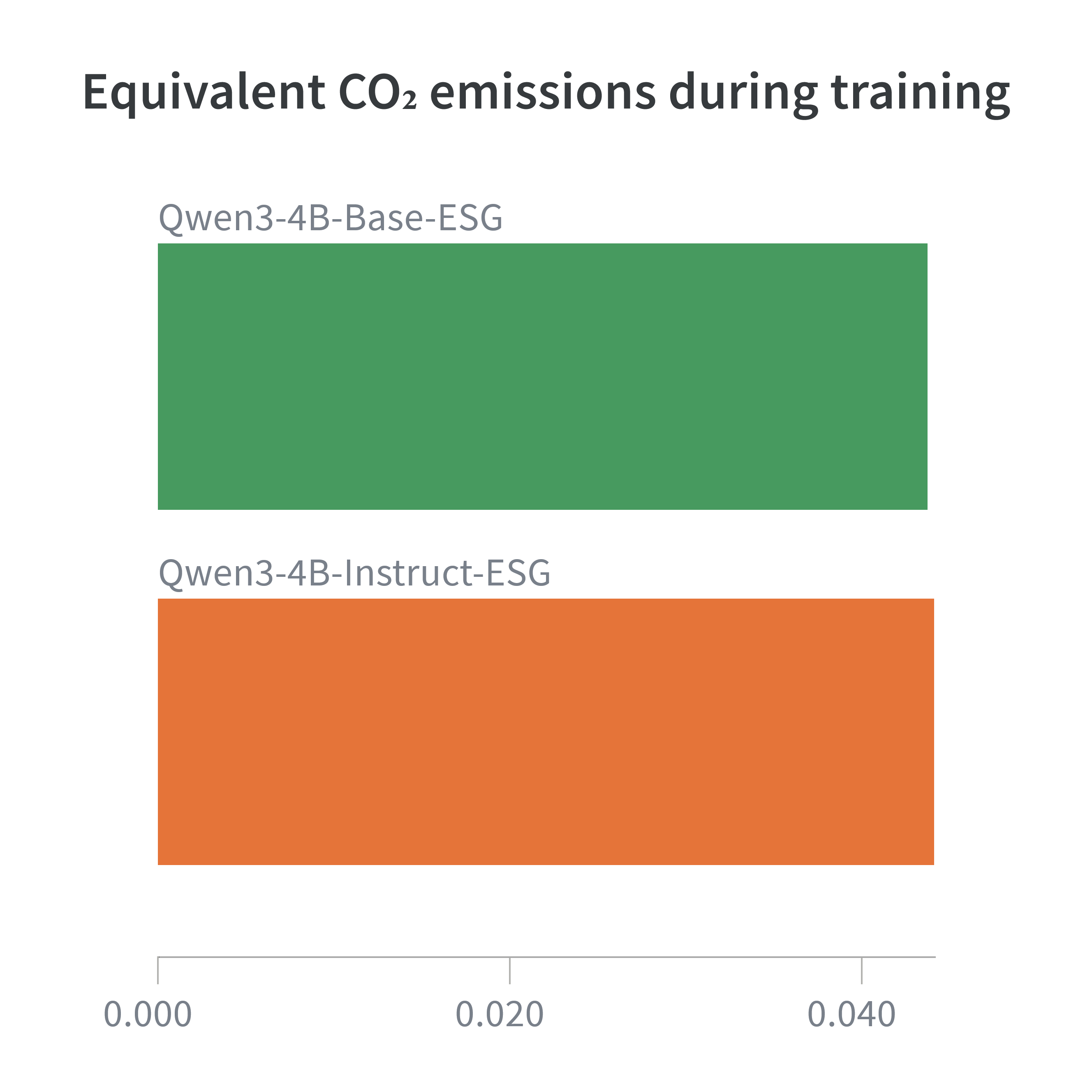}
      \caption{Training CO$_2$eq.}
      \label{fig:bottom}
    \end{subfigure}
    %\raggedleft 
    \caption{Loss and CO$_2$ impact.}
    \label{fig:training-regimes-metrics}
  %\end{flushright}
\end{figure}

Figure~\ref{fig:training-regimes-metrics} presents evidence from the training process. We observe comparable loss trajectories (Figure~\ref{fig:top}) across both regimes, with consistently lower loss when adapting the \emph{Base} model, which we attribute to its lack of prior instruction alignment and closer match to the distribution of the training texts. Figure~\ref{fig:bottom} also reports the estimated  CO$_2$-equivalent emissions computed with eco2AI~\citep{eco2AI}, which are similarly low across both approaches. The impact of IRM is negligible, as it involves only simple parameter addition and subtraction, and the measured emissions of approximately 0.04kg CO$_2$eq align with our commitment to environmentally responsible training.

\subsection{Evaluation Protocol}

Beyond assessing training performance, we evaluate model responses in an open-generation question answering (OQA) setting, following recent findings that purely discriminative evaluations often fail to capture the full extent of LLM knowledge and generative behavior~\citep{balepur-etal-2025-best}. In this setup, models must generate free-form answers rather than selecting predefined options, better reflecting real-world usage scenarios.

Accordingly, we employ a diverse set of OQA metrics. These include reference-based measures that quantify lexical overlap, such as F1, BLEU, METEOR, and ROUGE, as well as semantic-oriented metrics designed to capture meaning similarity beyond surface form, notably BERTScore~\citep{llms-qa-eval}. This combination allows us to assess both faithfulness and semantic adequacy of the generated responses.

In addition to response quality, and in line with responsible AI practices, we measure the environmental impact of inference using the \texttt{eco2AI} framework~\citep{eco2AI}, reporting energy consumption and equivalent CO$_2$ emissions. Finally, to analyze stylistic and accessibility aspects of model outputs, we evaluate readability using the \texttt{textstat} package, as a complementary signal related to the Social dimension of ESG, reflecting the clarity and accessibility of the generated responses. %Together, these complementary evaluation dimensions provide a comprehensive and principled assessment of ESG-oriented language model performance.

\section{Experimental Results}\label{sec:results}

In this section, we present and discuss the performance of the proposed models in a downstream ESG application. Section~\ref{sub:setup} details the evaluated experimental settings, while Section~\ref{sub:results} reports and analyzes the corresponding results.

\subsection{Experimental Setup}\label{sub:setup}

In addition to our proposed models, we include comparative evaluations with Llama-3~\citep{llama3}, Gemma-3~\citep{gemmateam2025gemma3technicalreport}, and the original Qwen-3~\citep{yang2025qwen3technicalreport}. Experiments are conducted under two evaluation regimes\footnote{All experiments were conducted using LlamaIndex~\citep{Liu_LlamaIndex_2022}.}.

\noindent\textbf{Zero-shot.} Models answer each ESG-QA test question directly, without examples or external context, probing knowledge encoded in the model parameters.

\noindent\textbf{Knowledge Base (KB).} Models are augmented with an embedding-based retrieval component ($\textit{top\_k}=3$) using \texttt{bge-small-en-v1.5}~\citep{bge_embedding}. The retrieval corpus comprises all ESG-QA contexts, approximating a realistic usage scenario.

Access to the KB was originally designed through model-initiated $\textit{tool\_calls}$ to a retrieval tool. However, during the experiments the fine-tuned models frequently produced malformed calls due to string-formatting errors. Therefore, for these models the KB was invoked programmatically by using the question as the retrieval query and inserting the retrieved passages into the context\footnote{Further details are provided in Appendix~\ref{appdx-A}.}. We acknowledge that this may introduce a mismatch in the evaluation setup, and future work will explore reliable approaches for model-initiated tool use while avoiding degradation caused by fine-tuning.

\subsection{Results}\label{sub:results}

First, Table~\ref{tab:gen_metrics_val} reports validation results for our training variants. All adapted models outperform the original Qwen-3-4B baselines, with the LoRA-tuned instruct model augmented with a KB achieving the best performance, and the original Qwen-3-Thinking model the worst. This pattern suggests that explicit reasoning capabilities, although beneficial for mathematical or logical tasks, are less effective for ESG-oriented dialogue.

Although the IRM-based models exhibited lower training loss, this advantage did not translate into superior downstream performance. This suggests that the data distribution and scale were insufficient to meaningfully degrade the instruction-tuned model's performance, which ultimately yielded stronger downstream results than the IRM variants.

\begin{table*}[ht!]
\renewcommand{\arraystretch}{1}
\setlength\tabcolsep{1.25mm}
\footnotesize
\caption{Generative performance on the validation set. Higher is better (\textuparrow). Best results are in \textbf{bold}, second-best are \underline{underlined}.}
\label{tab:gen_metrics_val}
\resizebox{\textwidth}{!}{%
\begin{tabular}{@{}l  ccccccccccc@{}}
\toprule
\multirow{2.5}{*}{\textbf{Mode/Model}}
  & \multirow{3}{*}{\makecell{\textbf{Rank}\\(Gen.)}}
  & \multirow{2}{*}{\makecell{\textbf{F1}}}
  & \multirow{2}{*}{\makecell{\textbf{METEOR}}}
  & \multirow{2}{*}{\makecell{\textbf{BLEU}}}
  & \multicolumn{4}{c}{\textbf{ROUGE}}
  & \multicolumn{3}{c}{\textbf{BERTScore}} \\
\cmidrule(lr){6-9}\cmidrule(lr){10-12}
  &
  &  &  & 
  & \textbf{R1} & \textbf{R2} & \textbf{RL} & \textbf{RLs}
  & \textbf{prec.} & \textbf{recall} & \textbf{f1} \\
\midrule

\textit{Baselines} \\

\makecell[l]{\textbf{Qwen3-4B-Instruct-2507}}  & 5
  & .170  & .301  & .034
  & .196  & .095  & .163  & .169
  & .827  & .893  & .859 \\

\makecell[l]{\textbf{Qwen3-4B-Thinking-2507}}  & 6
  & .145  & .241  & .015
  & .178  & .065  & .141  & .151
  & .804  & .885  & .842 \\

\midrule

\textit{This work} \\

\makecell[l]{\textbf{Qwen3-4B-Inst-ESG}}  & 3
  & .249  & .357  & .071
  & .279  & .150  & .237  & .238
  & .860  & .902  & .880 \\
\makecell[l]{\textbf{Qwen3-4B-Inst-ESG + KB}}  & 1
  & \textbf{.458}  & .595  & \textbf{.238}
  & \textbf{.504}  & \textbf{.366}  & \textbf{.454}  & \textbf{.454}
  & \textbf{.900}  & .933  & \textbf{.916} \\

\makecell[l]{\textbf{Qwen3-4B-IRM-ESG}}  & 4
  & .206  & .331  & .049
  & .236  & .119  & .197  & .200
  & .843  & .899  & .870 \\
\makecell[l]{\textbf{Qwen3-4B-IRM-ESG + KB}}  & 2
  & \underline{.436}  & \textbf{.597}  & \underline{.221}
  & \underline{.481}  & \underline{.348}  & \underline{.429}  & \underline{.430}
  & \underline{.894}  & \underline{.936}  & \underline{.914} \\

\bottomrule
\end{tabular}}
\end{table*}

\begin{table*}[ht!]
\renewcommand{\arraystretch}{1}
\setlength\tabcolsep{1.1mm}
\footnotesize
\caption{Generative performance on the test set. Higher is better (\textuparrow). Best results for the KB setup are shown in \textbf{bold}, while best zero-shot results are \underline{underlined}.}
\centering
%\resizebox{\textwidth}{!}{%
\label{tab:gen_metrics_test}

\begin{tabular}{@{}l ccccccccccc@{}}
\toprule
\multirow{2.5}{*}{\textbf{Mode/Model}}
  & \multirow{2.5}{*}{\makecell{\textbf{Rank}}}
  & \multirow{2.5}{*}{\makecell{\textbf{F1}}}
  & \multirow{2.5}{*}{\makecell{\textbf{METEOR}}}
  & \multirow{2.5}{*}{\makecell{\textbf{BLEU}}}
  & \multicolumn{4}{c}{\textbf{ROUGE}}
  & \multicolumn{3}{c}{\textbf{BERTScore}} \\
\cmidrule(lr){6-9}\cmidrule(lr){10-12}
  &
  &  &  &
  & \textbf{R1} & \textbf{R2} & \textbf{RL} & \textbf{RLs}
  & \textbf{prec.} & \textbf{recall} & \textbf{f1} \\
\midrule

% --- Family: Qwen-3-4B-esg-ep-3 ---
\textit{This work} \\
\makecell[l]{\textbf{Qwen3-4B-Inst-ESG}}  & 7
  & \underline{.246}  & \underline{.356}  & \underline{.069}
  & .276  & \underline{.146}  & .235  & .236
  & .861  & \underline{.902}  & \underline{.881} \\
\makecell[l]{\textbf{Qwen3-4B-Inst-ESG + KB}}  & 1
  & \textbf{.456}  & .593  & \textbf{.237}
  & \textbf{.502}  & \textbf{.365}  & \textbf{.452}  & \textbf{.452}
  & \textbf{.900}  & .933  & \textbf{.916} \\
\makecell[l]{\textbf{Qwen3-4B-IRM-ESG}}  & 10
  & .204  & .331  & .047
  & .233  & .115  & .194  & .197
  & .844  & .899  & .871 \\
\makecell[l]{\textbf{Qwen3-4B-IRM-ESG + KB}}  & 2
  & .437  & \textbf{.597}  & .222
  & .481  & .349  & .430  & .430
  & .894  & \textbf{.936}  & .914 \\
\midrule

% --- Family: Gemma-3 ---
\makecell[l]{\textbf{Gemma-3 4B}}  & 11
  & .191  & .256  & .030
  & .247  & .096  & .208  & .216
  & .849  & .887  & .867 \\
\makecell[l]{\textbf{Gemma-3 4B + KB}}  & 14
  & .085  & .098  & .015
  & .102  & .025  & .087  & .087
  & .819  & .847  & .832 \\
\makecell[l]{\textbf{Gemma-3 12B}}  & 8
  & .223  & .280  & .041
  & \underline{.280}  & .115  & \underline{.237}  & \underline{.242}
  & \underline{.866}  & .892  & .878 \\
\makecell[l]{\textbf{Gemma-3 12B + KB}}  & 3
  & .332  & .475  & .144
  & .371  & .242  & .325  & .325
  & .876  & .923  & .899 \\
\midrule

% --- Family: llama-3 ---
\makecell[l]{\textbf{Llama-3.1 8B}}  & 9
  & .211  & .333  & .043
  & .248  & .113  & .206  & .214
  & .847  & .903  & .874 \\
\makecell[l]{\textbf{Llama-3.1 8B + KB}}  & 5
  & .326  & .463  & .130
  & .368  & .228  & .316  & .316
  & .879  & .920  & .899 \\
\midrule

% --- Family: qwen-3 ---
\makecell[l]{\textbf{Qwen-3 4B}}  & 13
  & .154  & .255  & .018
  & .187  & .070  & .150  & .160
  & .812  & .888  & .848 \\
\makecell[l]{\textbf{Qwen-3 4B + KB}}  & 3
  & .340  & .446  & .138
  & .385  & .238  & .340  & .340
  & .886  & .916  & .900 \\
\makecell[l]{\textbf{Qwen-3 8B}}  & 12
  & .194  & .315  & .034
  & .228  & .103  & .186  & .193
  & .832  & .903  & .866 \\
\makecell[l]{\textbf{Qwen-3 8B + KB}}  & 6
  & .283  & .444  & .099
  & .323  & .195  & .274  & .278
  & .861  & .923  & .891 \\

\bottomrule
\end{tabular}
\end{table*}

Table~\ref{tab:gen_metrics_test} reports test-set results, including comparisons with other LLMs. Results under the KB setup indicate the Qwen3-4B-Inst-ESG variant as the best overall performer, closely followed by Qwen3-4B-IRM-ESG. Both outperform not only the original Qwen3 model but also all other models, including larger ones, suggesting that the combination of external knowledge and fine-tuned domain knowledge is beneficial. We again acknowledge that KB access strategies require further investigation, but current results show gains from incorporating external information after fine-tuning compared to the original model.

In the zero-shot setting, Qwen3-4B-Inst-ESG again outperforms its original counterpart and emerges as the strongest model in this configuration. The IRM variant also surpasses the original Qwen models, although it remains behind larger models such as Gemma-12B and Llama-8B in the zero-shot setting. Overall, the best-performing models rely on KB augmentation, indicating that while the adapted models internalize ESG knowledge and are well suited for direct use, augmented models remain more robust when external grounding is required.

\begin{figure}[!ht]
\centering
  %\begin{flushright} % left-align within ICLR template
    \begin{subfigure}[t]{0.5\textwidth}
      \centering
      \includegraphics[width=\linewidth]{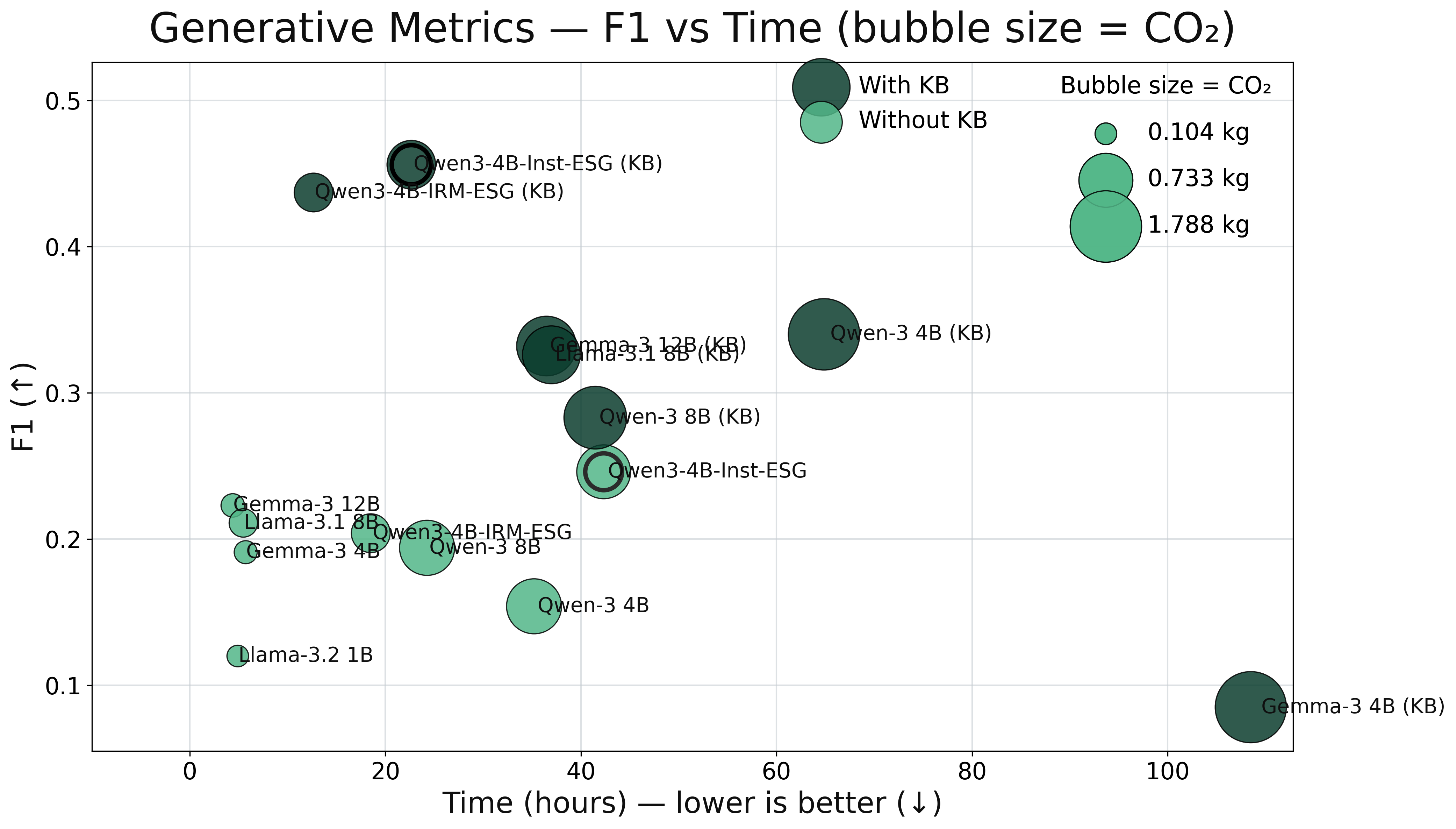}
      \caption{F1 / Time (Inference) / CO$_2$}
      \label{fig:F1}
    \end{subfigure}%\\[0.6em]
    \begin{subfigure}[t]{0.5\textwidth}
      \centering
      \includegraphics[width=\linewidth]{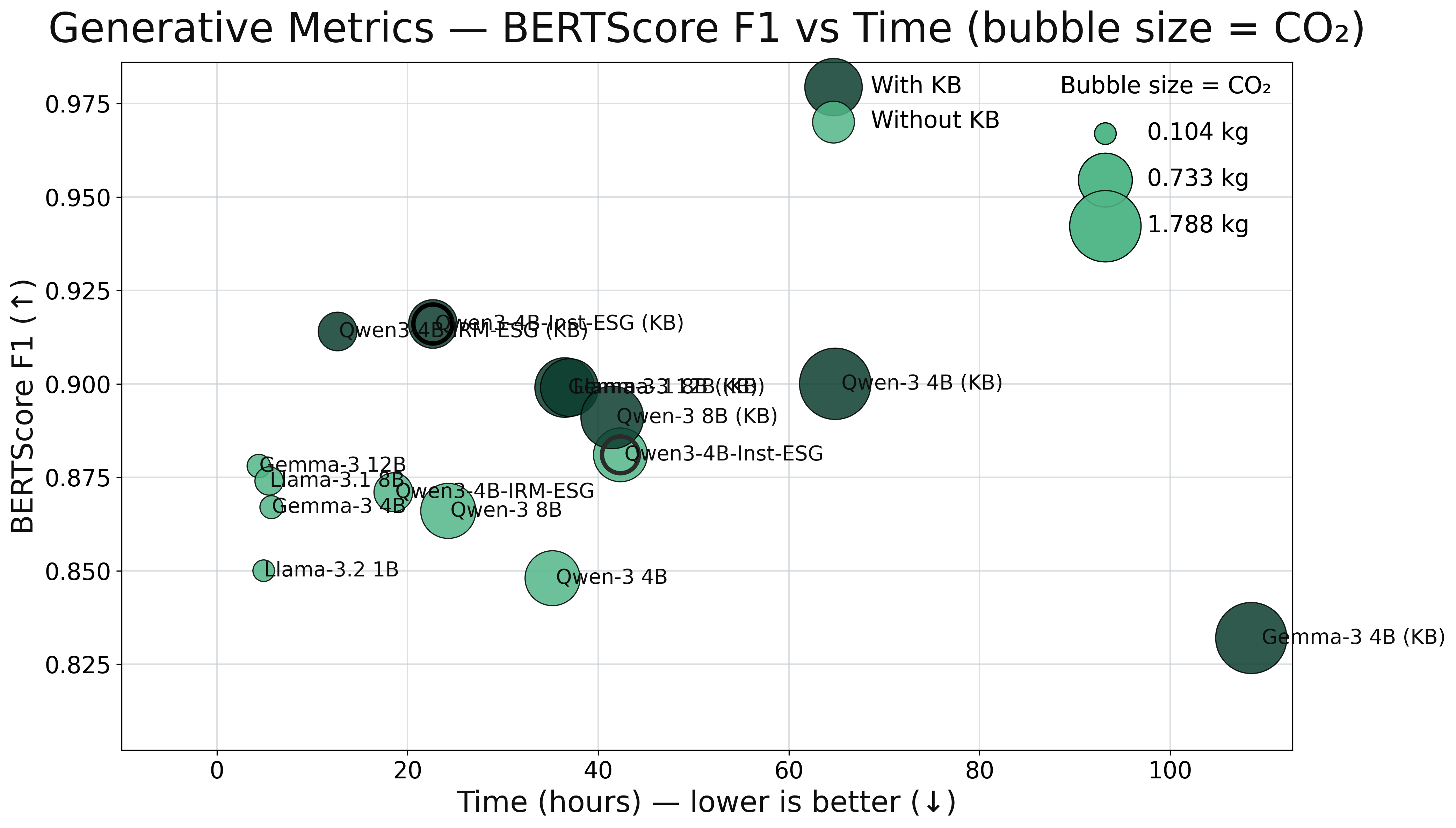}
      \caption{BERTScore F1 / Time (Inference) / CO$_2$}
      \label{fig:BERTSCORE}
    \end{subfigure}
    %\raggedleft 
    \caption{Trade-off between \textbf{inference impact} and generation quality on the test set.}
    \label{fig:gen_vs_co2}
  %\end{flushright}
\end{figure}

Figure~\ref{fig:gen_vs_co2} further reinforces these findings, complemented by inference-related computational indicators. The Qwen-3-4B-Inst-ESG model achieves the best performance in the zero-shot setting, albeit with increased inference time, likely due to higher output verbosity, as confirmed by manual inspection. When used with KB augmentation, inference time decreases, suggesting that external knowledge allows the model to better combine parametric knowledge and produce more direct responses. The IRM variant follows closely and is even more computationally efficient, confirming that IRM induces less degradation in augmented settings. Models augmented with a KB achieve the strongest generative scores overall; however, deploying and maintaining a KB is not always trivial or feasible in practice. In this context, Qwen-3-4B-ESG variants emerge as a compelling standalone option, while also presenting a lower CO$_2$ footprint.

\begin{table*}[ht!]
\renewcommand{\arraystretch}{1}
\setlength\tabcolsep{.75mm}
\scriptsize
\caption{Readability metrics on the test set.}
\centering
\label{tab:textstat_family}

\begin{tabular}{@{}l rrrrrrrrrrr@{}}
\toprule
\multirow{2.5}{*}{\textbf{Mode/Model}}
  & \multicolumn{1}{c}{\textbf{Direct. (\textuparrow)}}
  & \multicolumn{10}{c}{\textbf{Inverse (\textdownarrow)}} \\
\cmidrule(lr){2-2}\cmidrule(lr){3-12}
  & \textbf{FRE}
  & \textbf{FKG} & \textbf{Fog} & \textbf{SMOG} & \textbf{ARI} & \textbf{CLI}
  & \textbf{Linsear} & \textbf{DaleChall} & \textbf{DiffWords} & \textbf{Syllables} & \textbf{Sentences} \\
\midrule

\makecell[l]{\textbf{Reference}}  & 19.668
  & 14.905  & 18.847  & 15.847  & 15.954  & 16.581
  & 13.957  & 12.839  & 8.175  & 35.600  & 1.015 \\
\midrule

\textit{This work} \\
\makecell[l]{\textbf{Qwen3-4B-Inst-ESG}}  & 12.103
  & 17.097  & 21.444  & 17.633  & 19.830  & 18.028
  & 16.654  & 13.103  & 29.663  & 347.116  & 10.067 \\
  \makecell[l]{\textbf{Qwen3-4B-Inst-ESG + KB}}& 9.346  & 17.796  & 22.454  & 18.680  & 19.993  & 18.136  & 19.489  & 13.531  & 16.324  & 74.024  & \textbf{1.574} \\
  \makecell[l]{\textbf{Qwen3-4B-IRM-ESG}} & 9.965  & 17.107  & 21.507  & 18.085  & 18.982  & 18.232  & 17.487  & 13.280  & 35.766  & 180.298  & 4.326 \\
\makecell[l]{\textbf{Qwen3-4B-IRM-ESG + KB}} & 5.637  & 18.814  & 23.415  & 19.552  & 21.151  & 18.611  & 21.404  & 13.668  & 19.728  & 90.224  & 1.771 \\
\midrule

\makecell[l]{\textbf{Gemma-3 4B}}  & 14.722
  & \underline{14.376}  & \underline{18.083}  & \underline{14.204}  & \underline{15.752}  & 17.547
  & \underline{10.121}  & 13.625  & 17.261  & 72.975  & 2.738 \\
\makecell[l]{\textbf{Gemma-3 4B + KB}}  & \textbf{40.813}
  & \textbf{11.419}  & \textbf{13.768}  & \textbf{11.950}  & \textbf{14.751}  & \textbf{13.581}
  & \textbf{8.241}  & \textbf{11.604}  & \textbf{6.824}  & \textbf{47.025}  & 1.856 \\
\makecell[l]{\textbf{Gemma-3 12B}}  & 11.087
  & 15.314  & 19.156  & 15.169  & 17.170  & 18.286
  & 11.864  & 13.847  & \underline{13.894}  & \underline{57.710}  & \underline{2.026} \\
\makecell[l]{\textbf{Gemma-3 12B + KB}}  & 11.111
  & 17.214  & 21.386  & 18.093  & 18.942  & 17.929
  & 18.174  & 13.548  & 22.554  & 103.978  & 2.400 \\
\midrule

\makecell[l]{\textbf{Llama-3.1 8B}}  & 10.917
  & 17.136  & 21.252  & 17.766  & 18.889  & 17.643
  & 17.446  & 13.631  & 31.276  & 149.127  & 3.994 \\
\makecell[l]{\textbf{Llama-3.1 8B + KB}}  & 6.427
  & 19.382  & 23.471  & 19.662  & 21.897  & 17.718
  & 23.081  & 13.420  & 19.585  & 91.276  & 1.679 \\
\midrule

\makecell[l]{\textbf{Qwen-3 4B}}  & 13.860
  & 14.808  & 18.562  & 15.158  & 17.236  & 17.974
  & 11.175  & 14.129  & 37.562  & 181.045  & 7.071 \\
\makecell[l]{\textbf{Qwen-3 4B + KB}}  & -0.444
  & 19.207  & 23.380  & 18.670  & 21.553  & 18.670
  & 20.179  & 14.364  & 18.206  & 83.147  & 1.741 \\
\makecell[l]{\textbf{Qwen-3 8B}}  & 5.589
  & 16.790  & 20.963  & 17.041  & 19.112  & 19.348
  & 14.524  & 14.371  & 38.100  & 167.348  & 5.278 \\
\makecell[l]{\textbf{Qwen-3 8B + KB}}  & 4.440
  & 17.723  & 21.918  & 18.132  & 19.816  & 19.623
  & 17.264  & 14.104  & 30.860  & 136.227  & 3.385 \\

\bottomrule
\end{tabular}
\end{table*}

Table~\ref{tab:textstat_family} complements the analysis by reporting readability metrics for the generated responses. The results confirm that the Instruct-adapted model tends to produce more verbose outputs, with an average of approximately ten sentences per response. This pattern is not observed in the IRM variant or under KB augmentation, where the presence of additional context may reduce unnecessary generation. Models from the Gemma family achieve the highest readability scores, although they do not attain the strongest overall performance, highlighting a trade-off between readability and generative effectiveness that warrants further investigation in downstream tasks in this context.

Overall, the results indicate that our ESG Qwen-3 models provide a strong foundation for ESG-related scenarios, combining competitive performance with a modest inference carbon footprint.

\section{Conclusions}\label{sec:remarks}

This work introduces a family of ESG-oriented LLMs based on Qwen-3-4B, adapted using LoRA and IRM. The proposed models consistently outperform their original counterparts and demonstrate strong performance relative to models such as Llama-3 and Gemma-3. Importantly, the training pipeline is itself ESG-aware, relying on curated open data, environmentally responsible practices, and social considerations such as readability of generated outputs.

Despite these gains, some limitations remain. The adapted models exhibit reduced robustness in tool-based KB settings and increased verbosity, suggesting opportunities for future work involving preference alignment or reinforcement learning. Moreover, while we report extensive automatic evaluations, incorporating human assessments remains an important next step. Overall, this work provides a strong foundation for ESG-oriented language generation.

\section*{Acknowledgments}
This research was supported by CNPq (307088/2023-5); FAPERJ (SEI-260003/002930/2024, SEI-260003/000614/2023); and CAPES (Finance Code 001). The authors would also like to thank Instituto de Ciência e Tecnologia Itaú  (ICTi). Any opinions, findings, conclusions or recommendations expressed in this material are those of the authors and do not necessarily reflect the views of Itaú Unibanco and Instituto de Ciência e Tecnologia Itaú. This document does not constitute investment advice or any investment service. It is not and should not be deemed to be an offer to purchase or sell, or a solicitation of an offer to purchase or sell, or a recommendation to purchase or sell any securities or other financial instruments. All data used in this study comply with the Brazilian General Data Protection Law. ChatGPT (OpenAI) assisted with grammar, code and LaTeX, with all suggestions reviewed by the authors, who take full responsibility for the final content.

\bibliography{iclr2026_conference}
\bibliographystyle{iclr2026_conference}

\appendix
\section{External KB results with Qwen-ESG models}\label{appdx-A}

During our experiments leveraging a KB to extend model performance through external sources, we adopted a setup in which access to the KB was originally mediated by model-initiated $\textit{tool\_calls}$ to a retrieval component. However, we observed that our adapted Qwen-ESG models exhibited a substantial increase in projected execution time under this setup, with total runtime estimates reaching approximately 600 hours.

To better understand this behavior, we conducted an analysis of its potential causes. We found that the fine-tuned models frequently produced errors in function calling. The retrieval function, originally defined as \texttt{ESGRetriever}, was often incorrectly generated as \texttt{ESGRetriver}. This seemingly minor deviation resulted in a significantly higher number of function calls, with up to three times more calls required to answer a single query.

We conjecture that this degradation in function-calling accuracy is a side effect of the fine-tuning process, which negatively impacted the model's performance in agentic settings involving structured tool usage. Notably, qualitative inspection indicates that natural language generations remained well-formed and free of spelling errors, suggesting that the issue is localized to structured output generation rather than general language modeling.

To mitigate this limitation while preserving the benefits of KB integration, we explored two alternative configurations. In the first, denoted as \textit{e}KB, we removed the model's autonomy in deciding when to invoke the retrieval tool. Instead, the tool was called programmatically, using the input query directly as its key. In the second variant, denoted as \textit{n}KB, we simplified the function name to a more generic form (\texttt{search}) to reduce the likelihood of generation errors.

\begin{table*}[ht!]
\renewcommand{\arraystretch}{1}
\setlength\tabcolsep{1.25mm}
\footnotesize
\caption{Model performance in validation set -- generative metrics and inference consumption. Lower is better (\textdownarrow) for inference consumption, while higher is better (\textuparrow) for generative metrics. Best results are in \textbf{bold}, and second-best results are \underline{underlined}. \textit{e}KB denotes the programmatic external implementation of the knowledge base, whereas \textit{n}KB denotes the setup variant in which the name of the knowledge base tool is altered in the agentic setup.}
\label{tab:kb_sa}
\resizebox{\textwidth}{!}{%
\begin{tabular}{@{}l ccccccccccccccc@{}}
\toprule
\multirow{2.5}{*}{\textbf{Mode/Model}}
  & \multirow{3}{*}{\makecell{\textbf{Rank}\\(Gen.)}}
  & \multicolumn{4}{c}{\textbf{Inference Consumption}}
  & \multicolumn{10}{c}{\textbf{Generative Metrics}} \\
\cmidrule(lr){3-6}\cmidrule(lr){7-16}
  &
  & \multirow{2}{*}{\makecell{\textbf{Time}\\(h)}}
  & \multirow{2}{*}{\makecell{\textbf{Energy}\\(kWh)}}
  & \multirow{2}{*}{\makecell{\textbf{CO\(_2\)eq}\\(kg)}}
  & \multirow{2}{*}{\makecell{\textbf{vRAM}\\(GB)}}
  & \multirow{2}{*}{\makecell{\textbf{F1}}}
  & \multirow{2}{*}{\makecell{\textbf{METEOR}}}
  & \multirow{2}{*}{\makecell{\textbf{BLEU}}}
  & \multicolumn{4}{c}{\textbf{ROUGE}}
  & \multicolumn{3}{c}{\textbf{BERTScore}} \\
\cmidrule(lr){10-13}\cmidrule(lr){14-16}
  &
  &  &  &  &
  &  &  &
  & \textbf{R1} & \textbf{R2} & \textbf{RL} & \textbf{RLs}
  & \textbf{prec.} & \textbf{recall} & \textbf{f1} \\
\midrule

\textit{Instruct Variants} \\

\makecell[l]{\textbf{Qwen3-4B-Inst-ESG}} & 5
  & 24.948 & 3.761 & 0.425 & \underline{8.282}
  & \underline{.249} & \underline{.357} & \underline{.071}
  & \underline{.279} & \underline{.150} & \underline{.237} & \underline{.238}
  & \underline{.860} & \underline{.902} & \underline{.880} \\

\makecell[l]{\textbf{Qwen3-4B-Inst-ESG + \textit{e}KB}} & 1
  & \textbf{10.024} & \textbf{1.471} & \textbf{0.166} & 8.303
  & \textbf{.458} & \underline{.595} & \textbf{.238}
  & \textbf{.504} & \textbf{.366} & \textbf{.454} & \textbf{.454}
  & \textbf{.900} & .933 & \textbf{.916} \\

\makecell[l]{\textbf{Qwen3-4B-Inst-ESG + \textit{n}KB}} & 3
  & \underline{68.975} & \underline{11.923} & \underline{1.348} & \underline{14.244}
  & \textbf{.377} & \textbf{.547} & \textbf{.182}
  & \textbf{.414} & \textbf{.295} & \textbf{.367} & \textbf{.367}
  & \textbf{.888} & \textbf{.938} & \textbf{.912} \\

\midrule

\textit{IRM Variants} \\

\makecell[l]{\textbf{Qwen3-4B-IRM-ESG}} & 6
  & \underline{9.656} & \underline{1.392} & \underline{0.157} & \textbf{8.271}
  & .206 & .331 & .049
  & .236 & .119 & .197 & .200
  & .843 & .899 & .870 \\

\makecell[l]{\textbf{Qwen3-4B-IRM-ESG + \textit{e}KB}} & 2
  & 11.406 & 1.667 & 0.189 & 8.319
  & .436 & \textbf{.597} & .221
  & .481 & .348 & .429 & .430
  & .894 & \underline{.936} & .914 \\

\makecell[l]{\textbf{Qwen3-4B-IRM-ESG + \textit{n}KB}} & 4
  & 125.885 & 19.599 & 2.216 & 17.239
  & .262 & .403 & .120
  & .292 & .201 & .258 & .259
  & .846 & .920 & .881 \\

\bottomrule
\end{tabular}}
\end{table*}

Results in Table~\ref{tab:kb_sa} show that both approaches are effective. The \textit{e}KB configuration achieves the best overall performance while maintaining moderate execution time. The \textit{n}KB variant follows closely in performance, albeit with higher computational cost. Although the simplified function name reduces the frequency of call errors, we hypothesize that the increased latency in this setting is due to the agentic framework itself, which encourages multi-step reasoning in a ReAct-style paradigm. In contrast, the \textit{e}KB setup provides the retrieved context directly within the prompt, avoiding iterative tool interactions.

Overall, our findings suggest that fine-tuning can degrade a model's intrinsic ability to perform reliable function calls, particularly in agentic environments. Nevertheless, we demonstrate two practical and effective mitigation strategies. We believe that future work should further investigate this phenomenon, as well as evaluate additional models under similar configurations to enable a more comprehensive comparison.

\end{document}